# Effect of resonant magnetic perturbations on low collisionality discharges in MAST and a comparison with ASDEX Upgrade

A. Kirk[1], W. Suttrop[2], Yueqiang Liu[1], I.T. Chapman[1], P. Cahyna[3], T.Eich[2], C. Fuchs[2], C. Ham[1], J.R. Harrison[1], MW. Jakubowski[4], S. Pamela[1], M. Peterka[3], D. Ryan[1], S. Saarelma[1], R. Scannell[1], A.J. Thornton[1], M. Valovic[1], B. Sieglin[2], L. Barrera Orte[2], M. Willensdorfer[2], B. Kurzan[2], R. Fischer[2] and the MAST, ASDEX Upgrade and EUROfusion MST1 Teams

[1]CCFE, Culham Science Centre, Abingdon, Oxon, OX14 3DB, UK
[2]Max-Planck Institut für Plasmaphysik, Garching, Germany
[3]Institute of Plasma Physics AS CR v.v.i., IPP.CR, Prague, Czech Republic
[4]Max-Planck Institut für Plasmaphysik, Wendelsteinstr, Greifswald, Germany

## Abstract

Sustained ELM mitigation has been achieved on MAST and AUG using RMPs with a range of toroidal mode numbers over a wide region of low to medium collisionality discharges. The ELM energy loss and peak heat loads at the divertor targets have been reduced. The ELM mitigation phase is typically associated with a drop in plasma density and overall stored energy. In one particular scenario on MAST, by carefully adjusting the fuelling it has been possible to counteract the drop in density and to produce plasmas with mitigated ELMs, reduced peak divertor heat flux and with minimal degradation in pedestal height and confined energy. While the applied resonant magnetic perturbation field ($b^r_{res}$) can be a good indicator for the onset of ELM mitigation on MAST and AUG there are some cases where this is not the case and which clearly emphasise the need to take into account the plasma response to the applied perturbations. The plasma response calculations show that the increase in ELM frequency is correlated with the size of the edge peeling-tearing like response of the plasma and the distortions of the plasma boundary in the X-point region. In many cases the RMPs act to increase the frequency of type I ELMs, however, there are examples where the type I ELMs are suppressed and there is a transition to a small or type IV ELM-ing regime.

## 1.    *Introduction*

Type I ELMs are explosive events, which can eject large amounts of energy and particles from the confined region [1].  Extrapolation from present measurements suggest that the natural ELM frequency in ITER will vary from ~ 1 Hz for discharges with a plasma current $I_P$ = 15MA to  ~7Hz for $I_P$ = 5MA. Avoidance of both damage to Plasma Facing Components (PFC) and Tungsten (W) accumulation leads to a requirement that the ELM frequency is increased by a factor of ~3-40 over the natural ELM frequency as $I_P$ is increased from 5-15MA (see [2] and references therein).    Hence a mechanism is required to either increase the ELM frequency or to eliminate ELMs altogether accompanied by sufficient particle transport in order to avoid W accumulation. One such amelioration mechanism relies on perturbing the magnetic field in the edge plasma region, either leading to more frequent smaller ELMs (ELM mitigation) or ELM suppression.  This technique of Resonant Magnetic Perturbations (RMPs) has been employed to either mitigate or suppress type I ELMs on DIII-D [3][4], JET [5], ASDEX Upgrade [6], KSTAR [7] and MAST [8].

MAST and AUG are equipped with two rows of in vessel RMP coils (MAST: has 6 in the upper row and 12 in the lower row. AUG: 8 coils in both rows), which allow magnetic perturbations with a range of toroidal mode numbers (MAST: $n_{RMP}$=2, 3, 4, 6, AUG: $n_{RMP}$=1, 2, 4) to be applied. In this paper new results from MAST and ASDEX Upgrade at low pedestal top collisionality ($\nu^*_e$<1.0) will be presented.  In section 2 examples of the effect that applying RMPs with a range of toroidal mode numbers has on the ELM frequency will be reviewed. Section 3 discusses the advantages and disadvantages of ELM mitigation using RMPs. Section 4 describes an attempt on MAST to minimise the major disadvantage; namely, the density pump out, whilst maintaining the increase in ELM frequency. In section 5 the parameters determining the onset of ELM mitigation are studied, while section 6 examines what happens to the ELM type during mitigation.

## 2.      Examples of ELM mitigation on MAST and ASDEX Upgrade

ELM mitigation has been established on MAST in a range of plasmas using RMPs with toroidal mode numbers of $n_{RMP}$ = 3, 4 or 6 in both Lower Single Null (LSND) [9] and Connected Double Null (CDN) [10] magnetic configurations. Due to the ITER relevance of the LSND configuration, in the 2013 MAST campaign the discharge that was mostly used for the RMP experiments was based on a LSND H-mode plasma with a plasma current Ip = 400kA, which has an on-axis toroidal magnetic field $B_{T0}$ = 0.55T, an edge safety factor $q_{95}$ = 3.8 heated with 3.6MW of NBI power. Typical values at the pedestal top are; electron density $n_e^{ped}$ = 1.0-4.0x10$^{19}$ m$^{-3}$, the electron temperature $T_e^{ped}$ = 150-250 eV corresponding to electron collisionality ($\nu_e^*$) = 0.3-0.8, where the collisionality is calculated following reference [11] as:

$$\nu_e^* = 6.921.10^{-18} \frac{Rqn_e Z_{eff} \ln \Lambda_e}{\varepsilon^{3/2} T_e^2}$$

where R is the major radius in m, $q_{95}$ is the safety factor at 95% of flux surface and where $\varepsilon$ is the inverse aspect ratio. $Z_{eff}$ is the effective ion charge, $n_e$ the electron density in m$^{-3}$ and $T_e$ the temperature in eV evaluated at the top of the pedestal. $\ln\Lambda_e$ is the Coulomb logarithm defined by $\ln \Lambda_e = 31.3 - \ln(\sqrt{n_e} / T_e)$.

The previous LSND studies on MAST [9] used a reference discharge with Ip = 600kA, which had a lower central q.  Throughout the H-mode period, this discharge had sawtooth activity; this coupled to the edge and resulted in an ELM simultaneous with the sawtooth crash.  The lower $I_P$ discharge presented here has no sawtooth activity before 0.5 s, which allows a period of H-mode unaffected by core activity.   Figure 1 shows an example of the effect that the application of RMPs with toroidal mode number $n_{RMP}$ = 2, 3, 4, 6 has on this discharge.  The change in ELM behaviour can be seen from the changes in the $D_\alpha$ emission when the current in the in-vessel coils (Figure 1a) is applied for different RMP configurations. Also shown is the enhanced particle transport, or so-called density pump-out (Figure 1b), which occurs when RMPs are applied as well as the braking of the toroidal rotation velocity at the top of the pedestal (Figure 1c). For the $n_{RMP}$=2 and 3 perturbations the rotation braking and density pump out is so large that it leads to a back transition to L-mode. Studies on the effect that RMPs applied before the L-H transition have on the power required to access H-mode have shown a non-monotonic increase with $n_{RMP}$ [12], with the least impact on the power required to access H-mode occurring for $n_{RMP}$ = 4 [13].

On ASDEX Upgrade complete suppression of type I ELMs has been achieved at high density [6] but there have been fewer studies of discharges at low collisionality. The AUG B-coil set [6] can apply RMPs with a range of toroidal mode numbers $n_{RMP}$=1, 2 and 4. Figure 2 shows examples of the effect that the application of RMPs with toroidal mode number $n_{RMP}$ = 2, 4 have on a LSND H-mode plasma with a plasma current Ip = 800kA, an on-axis toroidal magnetic field $B_{T0}$ = 1.8T and an edge safety factor $q_{95}$ = 3.6 heated with 6 MW of NBI power and up to 3.4 MW of ECRH. Typical values at the pedestal top are; electron density $n_e^{ped}$ = 1.8-4.2x$10^{19}$ m$^{-3}$, the electron temperature $T_e^{ped}$ = 800-1200 eV corresponding to an electron collisionality ($\nu_e^*$) = 0.03-0.4. After the coil current (Figure 2a) reaches a certain threshold there is a density pump out (Figure 2b) and a large reduction in the ELM size (Figure 2d and e). Figure 3 shows a zoom in of a 5ms period during the mitigation stage for shot 31128 with the application of the RMPs in an $n_{RMP}$ = 2 configuration. As can been seen this period is characterised by very small ELMs with an ELM frequency of up to 800 Hz. Similar to what is observed on MAST, the toroidal rotation at the top of the pedestal (Figure 2c) is also seen to drop although in the case of AUG the deceleration is less severe. ELM mitigation has also been achieved using $n_{RMP}$ = 1 [14].

### 3. Pros and cons of ELM mitigation

Figure 4 a) and c) show the ELM frequency ($f_{ELM}$) versus the energy loss per ELM ($\Delta W_{ELM}$), derived from the change in plasma stored energy calculated from equilibrium reconstruction, for the natural and mitigated ELMs in MAST and AUG respectively. The application of the RMPs produces an increase in $f_{ELM}$ and corresponding decrease in $\Delta W_{ELM}$ consistent with $f_{ELM}.\Delta W_{ELM}$= const (represented by the dashed curves on the figures). In the case of AUG the increase in ELM frequency is up to a factor of 10 and the ELM size drops to less than 2 kJ per ELM, which is at the limit of the resolution of the calculated plasma stored energy.

In order to avoid damage to in-vessel components in future devices, such as ITER, the peak heat flux density ($q_{peak}$) at the divertor during the ELM is more important than $\Delta W_{ELM}$. The divertor heat fluxes on MAST and AUG have been measured using infrared thermography. Figure 4b and d) show $q_{peak}$ at the outer target for MAST and inner target for AUG as a function of $\Delta W_{ELM}$. The reason for the difference in the choice of divertor target monitored is due to the fact that on MAST the majority of the ELM energy arrives at the outer target [15] while on AUG the majority of the energy arrives at the inner target [16]. The increase in ELM frequency and decrease in $\Delta W_{ELM}$ does lead to reduced heat fluxes at the target; the decrease is not linear due to the target wetted area [17] getting smaller as the ELM size is reduced. These results are more favourable than the results obtained in JET with a carbon wall [19], where it was found that the reduction in wetted area was such that the peak heat flux was not mitigated. The MAST results show that for the same ELM frequency, the reduction in the ELM energy loss and peak heat load on the divertor plates is approximately the same for all RMP configurations.

While the reduction in peak heat flux is important the more relevant quantity for material limits is the heat flux factor ($\eta_{ELM}$), which is defined as the energy of the ELM to the divertor normalised by the wetted area and the square root of the ELM duration [17] For the MAST cases discussed here $\eta_{ELM}$ decreases from an average value of 0.31 MJm$^{-2}$s$^{-0.5}$ for ELMs with $\Delta W_{ELM} = 8$ kJ to 0.17 MJm$^{-2}$s$^{-0.5}$ for $\Delta W_{ELM} = 4$kJ [20]. Similarly in the cases of the AUG discharges $\eta_{ELM}$ decreases from 0.35 MJm$^{-2}$s$^{-0.5}$ for $\Delta W_{ELM} = 40$ kJ to 0.09 MJm$^{-2}$s$^{-0.5}$ for $\Delta W_{ELM} = 10$ kJ. In both cases this represents a significant reduction in the impact that ELMs have on the material surfaces.

Figure 5 shows the effect that RMPs in an $n_{RMP} = 6$ configuration in MAST and an $n_{RMP} = 2$ configuration in AUG have on the radial profiles of the electron temperature and density. The profiles were obtained in the last 10 % of the ELM cycle. The RMPs lead to a similar reduction in the density profile and leave the electron temperature effectively unchanged. This reduction in density at constant electron temperature means that the overall plasma stored energy is reduced by between 20 and 30 %. Hence there is a price to pay in confinement for the reduced target heat loads and therefore it is important to find ways of minimising this density pump out.

**4.      Minimising the effects of the density pump out on MAST**

Figure 6d and e show the $D_\alpha$ traces for a pair of shots on MAST that have $I_{ELM}= 0$ or 4 kAt with the RMPs in an $n_{RMP} = 6$ configuration, where there is no gas fuelling during the H-mode period. As the RMPs are applied the type I ELM frequency increases from the natural value of $f_{ELM} = 50$ Hz for $I_{ELM} = 0$ to $f_{ELM} = 150$ Hz for $I_{ELM}=4$ kAt and the density decreases, eventually leading to a back transition to L-mode. This is because the power to remain in H-mode ($P_{L-H}$) has a minimum value at a certain density and below this density the power required to stay in H-mode rises sharply [18]. Figure 6f shows a discharge in which this minimum density has been avoided by increasing the density feedback set point in the plasma control system. As the pump out begins the gas fuelling increases (Figure 6c) and the drop in the density is arrested leading to sustained ELM mitigation. To demonstrate the strong correlation on MAST between density transport and the size of the applied perturbation the discharge has been repeated with a higher RMP field strength using $I_{ELM}=4.8$ kAt. For this case the gas refuelling rate required to maintain the same density increases (Figure 6c and g).

While it was possible to maintain the plasma density using this technique, it was not possible to refuel the discharge to the non-RMP density without the gas refuelling rate being so high that it substantially degraded the temperature pedestal. Instead an alternative method was used whereby a constant gas puff was used combined with a slow ramp of the RMP field. The gas fuelling rate was adjusted to compensate the pump out. Figure 7 shows a pair of discharges without and with RMPs in an $n_{RMP} = 6$ configuration with a constant gas puff fuelling rate in the shot with the RMPs applied. The ELM frequency increases from ~60 Hz to ~230 Hz while the line average density is kept constant. To investigate the effect of the gas puff rate on the natural ELM frequency the RMP off shot was repeated with the gas puff rate the same as in the RMP on shot. The density increased by less than 10% and the natural ELM frequency increased to ~ 80 Hz.

Figure 8a and b show that the radial profiles of the electron density and temperature for these two discharges are similar. Figure 8d and e show that the density lost per ELM and the energy loss per ELM is much smaller in the mitigated case ($\Delta W_{ELM}$ decreases from 8kJ to 2kJ and $q_{peak}$ decreases from 10 MWm$^{-2}$ to 3 MWm$^{-2}$). Figure 8e shows that the overall stored energy is almost the same. Hence for this discharge it has been possible to increase the ELM frequency and reduce the ELM size by a factor of 4 while maintaining the same overall confinement and pedestal pressure.

These results also demonstrate that the pedestal pressure alone is not responsible for determining $\Delta W_{ELM}$ and $q_{peak}$. Figure 9 compares the difference in the density pedestal profiles before and after a natural and mitigated ELM (i.e. $\Delta n_e(R) = n_e^{before\ ELM}(R) - n_e^{after\ ELM}(R)$), which shows that while the peak change in density is similar the radial extent of the losses are much reduced in the mitigated ELM. Hence it is the ELM affected area that is reduced in the mitigated cases.

Previous studies on MAST [2] have found that the pressure pedestal ($P^{ped}$) evolves continuously in a similar way between natural and mitigated ELMs; however, for the mitigated ELMs the ELM is triggered earlier in the ELM cycle at a lower value of $P^{ped}$, reflecting the increased ELM frequency. For the shots presented in this paper the inter-ELM pedestal evolution is again similar for natural and mitigated ELMs but in the natural ELM cycle the pressure pedestal spends a large fraction of time near to a saturated value (see Figure 10). Therefore, it is possible to increase the ELM frequency without substantially degrading the pedestal by arranging that the mitigated ELMs are triggered near to the point at which the maximum pressure pedestal value is first obtained. This appears to be the case for the mitigated ELMs presented here (Figure 10). It is likely that if $f_{ELM}$ was increased further, by for example increasing the perturbation strength, then the peak $P^{ped}$ value obtained would be reduced and this would affect the edge and possibly the core confinement.

Refuelling studies have also been performed in these discharges using the injection of frozen deuterium pellets from the high field side [21], in this case it has been possible to refuel to densities higher than in the natural ELM case while still keeping the ELM mitigation. In these cases there is an overall drop in confinement of ~ 10 % but they do demonstrate that pellet fuelling is compatible with ELM mitigation.

**5.    Parameters determining the onset of ELM mitigation**

Previous studies on MAST have shown that the RMPs give rise to perturbations of the plasma shape, with lobe structures forming due to the tangled magnetic fields near the X-point [22], and corrugations of the plasma boundary at the mid-plane [23]. The X-point lobe length increases linearly with the resonant components of the applied field ($b^r_{res}$) when above a threshold value, with higher $n_{RMP}$ giving rise to longer lobes for the same $b^r_{res}$ [24]. Similarly, the mid-plane displacement increases with $b^r_{res}$, though the corrugation amplitude is less dependent upon the RMP configuration [23]. The mitigated ELM frequency increases with $b^r_{res}$ calculated in the vacuum approximation provided it is above a critical threshold [9][10]. This threshold value depends on the mode number of the RMP, with higher $n_{RMP}$ having a larger critical value.  Although a similar dependence is observed in the LSND discharges with $I_P$=400 and 600 kA the thresholds are different (see Figure 11). These calculations have been performed in the vacuum approximation and assuming that a single dominant toroidal mode number is responsible for the effects.

A similar trend of increasing ELM frequency with $b^r_{res}$ has been observed on AUG, However, as can be seen from Figure 12 there are some clear outliers.  On AUG the pitch angle of the applied field can be changed in an $n_{RMP}$=2 configuration by exploiting the fact that there are more coils than required to produce the toroidal mode number. By adjusting the current in the B-coils the phase difference ($\Delta\phi$) between the field patterns in the upper and lower row can be modified. The two open circles in Figure 12 come from two identical discharges in which the phase of the perturbation has been changed from $\Delta\phi = 90^\circ$ to $180^\circ$ (see Figure 13). Both coil configurations have a similar effect on the plasma in terms of density pump out and increase in ELM frequency.

The vacuum modelling for these shots is shown as the solid symbols in Figure 14, which shows that if the size of $b^r_{res}$ at the plasma edge is the most important quantity then the 90° phasing should have the largest effect on the plasma. Calculations have been performed using the MARS-F code, which is a linear single fluid resistive MHD code that combines the plasma response with the vacuum perturbations, including screening effects due to toroidal rotation [25]. The calculations use the experimental profiles of density, temperature and toroidal rotation as input and realistic values of resistivity, characterised by the Lundquist number (S) which varies from $\sim 10^8$ in the core to $\sim 10^6$ in the pedestal region (the radial profile of the resistivity is assumed proportional to $T_e^{-3/2}$). The resistive plasma response significantly reduces the field amplitude near rational surfaces. It reduces the resonant component of the field by more than an order of magnitude for $\Psi_{pol}^{\frac{1}{2}} < 0.97$ (see open symbols in Figure 14) and results in the 90° and 180° cases having very similar values of $b^r_{res}$ at the plasma edge, which may then explain why both have a similar effect on the ELM frequency.

Figure 15 shows the $b^r_{res}$ at the q=5 surface as a function of the phase angle between the upper and lower row of coils calculated in the vacuum approximation and taking into account the plasma response. The peak value of $b^r_{res}$ at the q=5 surface occurs at $\Delta\phi \sim 60°$ in the vacuum approximation and $\Delta\phi \sim 120°$ when the plasma response is taken into account i.e. an offset of $\sim 60°$. The offset arises due to the resistive plasma response near to the plasma edge [26]. In the core of the plasma, where the response is closer to being ideal, the offset is $\sim 90°$. A similar shift in the optimum pitch-resonant alignment has been previously observed both experimentally and in modelling ELM suppression experiments with n=2 magnetic perturbations on DIII-D [27].

Whilst a full scan in $\Delta\phi$ has not been performed on this shot, one has been performed on a similar low $\nu^*$ shot that has a slightly lower toroidal field on-axis ($B_T$=1.79 T) and shape. These changes lead to a different value of $\Delta\phi$ for which the optimum pitch-resonant alignment occurs. The effect of experimentally varying $\Delta\phi$, at constant $I_{Bcoil}$, from +100° to -100° is shown in Figure 16. The ELM frequency is observed to increase and the density reduce as soon as the RMPs are turned on with $\Delta\phi = 100°$. As $\Delta\phi$ is reduced the level of mitigation gradually reduces until $f_{ELM}$ returns to the RMP off value at $\Delta\phi \sim -10°$.

The ELM frequency continues to reduce until a classic ELM free period is encountered, at which point the density rises rapidly leading to a steady type I ELM-ing regime. The ELM frequency as a function of $\Delta\phi$ is shown in Figure 17. MARS-F calculations have been performed for this discharge as a function of $\Delta\phi$. Figure 18 shows the $b^r_{res}$ at the q=5 surface as a function of $\Delta\phi$ calculated in the vacuum approximation, which peaks at $\Delta\phi = 30°$ and taking into account the plasma response, which peaks at $\Delta\phi = 90°$. The difference in the location of the peak positions is again ~60°. As can be seen the trend of $b^r_{res}$ with $\Delta\phi$ calculated taking into account the plasma response is in good agreement with the observed change in ELM frequency up to the point that discharge enters into an ELM free period ($\Delta\phi \sim$ -85°).

The plasma response leads to plasma displacements normal to the flux surfaces [28][29]. The radial profiles of the poloidal (*m*) Fourier harmonics (in a PEST-like straight line coordinate system [30]) often peak at low *m* in the plasma core (referred to as a core kink component) and high *m* near the plasma edge (referred to as an edge peeling-tearing component) [28]. Whether a core kink component, or the edge peeling-tearing component, or both appear in the plasma response, depends on the plasma equilibrium and the coil configuration. Figure 19 shows an example of the radial profiles of the poloidal Fourier harmonics in which both the core kink and edge peeling-tearing components are both present in the plasma response calculation. Figure 20a shows the size of the maximum displacement ($\zeta_m$) for the core kink and edge peeling-tearing like response as a function of $\Delta\phi$. The core kink response is effectively symmetric around $\Delta\phi = 0$, whereas as the edge peeling-tearing like response is similar to the trend observed experimentally for $f_{ELM}$.

These core kink and edge peeling-tearing like responses lead to a deformation of the plasma surface, which varies with poloidal and toroidal location. It has previously been observed on MAST that there is a clear correlation between the location of the maximum of the amplitude of the normal component of the plasma displacement at the plasma surface and the effect of the RMPs on the plasma [28]. In these studies it was observed that a density pump out in L-mode or ELM mitigation in H-mode only occurred when the displacement at the X-point was larger than the displacement at the mid-plane combined with the requirement that the X-point displacement was larger than a critical value [9]. Figure 20b shows the peak mid-plane and X-point displacements ($\zeta_{edge}$) as a function of $\Delta\phi$. There is a clear correlation between the core kink like response and the mid-plane displacement and a similar correlation between the edge peeling-tearing like response and the X-point peaking, which is also correlated with the experimentally observed ELM frequency.

To enable a more systematic study of the effect of the various components on the ELM frequency, the experimentally measured ELM frequency and the parameters discussed above have been mapped onto the same $\Delta\phi$ basis and then plotted against each other. The data after the ELM free period (i.e. $\Delta\phi < -85°$) have been excluded from this mapping. Given the uncertainties in the modelling etc. it is interesting to see from Figure 21a that above a threshold value of $b^r_{res} \sim 0.05 \times 10^{-3}$ the increase in $f_{ELM}$ scales almost linearly with $b^r_{res}$ calculated taking into account the plasma response. A similar linear scaling above a threshold value has also been observed on MAST where the threshold was toroidal mode number dependent ($0.08 \times 10^{-3}$ for $n_{RMP} = 4$ and $0.15 \times 10^{-3}$ for $n_{RMP} = 6$) [9]. Figure 21b shows $f_{ELM}$ versus the peak amplitude of the core kink and edge peeling-tearing like responses, which confirms the good correlation with the edge peeling-tearing like component. Figure 21c shows $f_{ELM}$ versus the plasma surface displacement. Above a threshold of $\sim 1.5$mm $f_{ELM}$ increases linearly at first with $\zeta_{edge}$. In is interesting to note that this threshold of 1.5 mm is very similar to the threshold obtained on MAST for both $n_{RMP} = 4$ and 6 RMP configurations [9].

All these simulations performed to date have assumed that the effects are due to a dominant toroidal mode number. The need to consider how all the toroidal mode numbers may couple to produce the effect on $f_{ELM}$ can be seen in experiments performed on MAST to simulate the effect of the failure of a set of coils. Figure 22d shows a LSND discharge on MAST, similar to that shown in Figure 6 and described above, in which the ELMs are mitigated with $I_{ELM} = 4$ kAt using an $n_{RMP} = 6$ configuration of the RMP coils. In subsequent shots only 9 of the 12 coils were powered and the current in ELM coils was increased until at $I_{ELM} = 4.8$ kAt a similar level of ELM mitigation was obtained (Figure 22d). Although the coil current has increased, the $n_{RMP} = 6$ resonant field component has reduced by 20 % (see Figure 23a). Figure 23b shows the Fourier decomposition of the radial field produced by the coils at the location of the last closed flux surface at the low field side mid-plane. In the case of 12 coils being used the field has effectively a pure $n_{RMP} = 6$ harmonic. However, in the 9 coil case, there are significant contributions to toroidal harmonics $n_{RMP}=1$ through to $n_{RMP}=9$. The fact that a similar level of ELM mitigation is achieved suggests that mixing of different toroidal harmonics may be important.

## 6. What happens to the ELM type during mitigation?

Previous studies of the filament structures observed during ELMs on MAST have shown that the natural and mitigated stages have similar characteristics [9]. These studies have been repeated for the $I_P=400$ kA discharges described in this paper using a new, improved resolution camera. The images obtained using a 3.5 μs exposure have been analysed during the rise time of the mid-plane $D_\alpha$ signal for natural ($I_{ELM} = 0$) and mitigated ELMs obtained using an $n_{RMP}=4$ configuration with $I_{ELM}=5.2$ kAt. The average ELM energy loss in the two cases is 7.0 and 1.5 kJ respectively. The mean separation in the toroidal angle between the filament locations is used to derive an effective toroidal mode number, shown in Figure 24a, which has a mean value of ~17 for both natural and mitigated ELMs. The toroidal width of the filaments has been determined from the width of the intensity distribution; Figure 24b, which shows that the filament widths, at least in the toroidal direction, are not affected by the application of the RMPs. Whilst the density pedestal reduces at constant temperature pedestal the pedestal height characteristics remain in the region typically associated with type I ELMs on MAST [31]. So in the case of the MAST discharges described in this paper to date, the filament structures and pedestal characteristics suggest that the ELM character does not change i.e. they remain type I ELMs.

A high frequency ELM-ing regime has also been observed on MAST both naturally [32] and with the application of RMPs [10] in a so called scenario 4 discharge, which is based around a neutral beam heated plasma with $I_P = 750$ kA, $B_T = 0.55$ T in a connected double null magnetic configuration with $q_{95}= 5.4$. For these shots the pedestal top values are; the electron density $n_e^{ped} = 2.0\text{-}4.0\text{x}10^{19}$ m$^{-3}$ and the electron temperature $T_e^{ped} = 150\text{-}250$ eV corresponding to an electron collisionality $v_e^* = 0.3\text{-}0.8$. Previous studies of this discharge without the application of RMPs have shown that at low density the ELMs move from being type I to a small ELM-ing regime where the ELMs have been identified as being type IV in nature (often referred to as the low collisionality branch of type III ELMs) [31].

Figure 25c shows the target $D_\alpha$ time trace for the baseline type I ELM-ing shot that does not have the RMPs applied, which has an ELM frequency of ~ 100 Hz and an approximately constant line average density (Figure 25b). This scenario has no gas puffing from 180ms and the refuelling is due to recycling from the targets and the residual neutral density in the vessel. Figure 25d shows the $D_\alpha$ trace for a shot where the application of the RMPs results in a density pump out (Figure 25b) shortly after the current in the ELM coils reaches the maximum value of 5.6kAt. Following the reduction of the density, the plasma attains a stable state with high frequency (~2000 Hz) small ELMs. The density and electron temperature profiles before, during and after the density pump out event have been measured by a Thomson scattering system and show that the pedestal characteristics change from a region typically associated with type I ELMs to one associated with naturally occurring type IV ELMs (Figure 25e) [31].

The filament characteristics during the ELMs have been investigated in the type I ELM period and during the small ELMs. Figure 26a shows the effective toroidal mode number obtained, which in the type I ELM-ing has a mean value of ~18. However, during the type IV ELM period produced either naturally or by the application of RMPs the mode number increases to a mean value of ~24. Finally in the case of naturally occurring type IV ELMs the ELM frequency is found to increase as the pedestal density decreases (Figure 26b), which is opposite to what is normally observed for type I ELMs [32].

In the case of AUG no information is available on the filaments characteristics. However, the change in the pedestal values and the variation of ELM frequency with pedestal density may indicate a change in ELM type. Figure 27a shows that in the mitigated stage the temperature pedestal remain constant but the density pedestal reduces such that the pedestal characteristics of the mitigated ELMs are in a region where high frequency ELMs occur naturally on some other devices [32]. In addition, the size of the mitigation increases as the density pedestal is reduced (see Figure 27b). Both characteristics suggest that this may be a transition to type IV ELMs. The points in Figure 27b that have $f_{mit}/f_{nat} < 1$ occur in the discharges where a pump in is observed and the density rises above the natural pedestal density of $\sim 3.5 \times 10^{19}$ m$^{-3}$.

The only disadvantage of the type IV ELM mitigated regime on MAST is that refuelling this regime leads to a decrease in $f_{ELM}$ and increase in $\Delta W_{ELM}$ or an eventual transition back to large type I ELMs [10]. The effect of refuelling is something that will need to be studied in future experiments on AUG.

## 7.    Summary and discussion

Sustained ELM mitigation has been achieved on MAST and AUG using RMPs with a range of toroidal mode numbers in a wide region of low to medium collisionality discharges. The ELM mitigation phase on both devices results in smaller ELM energy loss and reduced peak heat loads at the divertor targets. The ELM mitigation phase is typically associated with a drop in plasma density and overall stored energy. However, on MAST, in one particular plasma scenario where the pedestal pressure remains at a saturated value for a large fraction of the natural ELM cycle, by carefully adjusting either the gas or pellet fuelling, it has been possible to produce plasmas with mitigated ELMs, reduced peak divertor heat flux and with minimal degradation in pedestal height and confined energy.

On MAST above a threshold value in the applied perturbation field ($b^r_{res}$) there is a linear increase in normalised ELM frequency ($f_{ELM}$) with $b^r_{res}$. Experimentally it has been found that both the lobes produced near the X-point [24] and the mid-plane corrugations also increase linearly with the size of $b^r_{res}$[23]. These deformations to the plasma boundary have been replicated by modelling, which shows that they can strongly influence the peeling-ballooning stability boundary and hence lead to an increase in $f_{ELM}$ [33]. On AUG, in a large number of cases an increase of $f_{ELM}$ with $b^r_{res}$ is also observed. However, unlike in MAST, there are examples where this is not the case. These cases clearly demonstrate the need to take into account the plasma response to the applied perturbations. When this is done it is found that the increase in the ELM frequency is correlated with the edge peeling-tearing like response of the plasma and the edge displacement of the plasma near to the X-point.

An analysis of the filament structures observed during the ELMs on MAST suggest that in most cases the ELMs remain type I in nature, however, there is one scenario on MAST where a small ELM regime exists in which the pedestal and filament characteristics are those more typically associated with type IV ELMs. Such a suppression of type I ELMs and transition to a type IV ELM-ing regime may also be occurring in AUG. In reference [2] the regions of operational space for which type I ELMs have been suppressed or mitigated was represented in a plot of pedestal collisionality ($v^*_e$) versus line average density expressed as a fraction of the Greenwald number ($n_e/n_{GW}$). Figure 28 shows the updated version of these plots including the new results from MAST and AUG presented in this paper. The MAST results, for the main part fall in the ELM mitigation plot, with the results on the transition to type IV ELMs included in the type I ELM suppression plot. The new AUG results have been included in both plots since there are periods in which the Type I ELM frequency is increased as well as the region where there is a possible transition to type IV ELMs.

## Acknowledgement

This work has been carried out within the framework of the EUROfusion Consortium and has received funding from the European Union's Horizon 2020 research and innovation programme under grant agreement number 633053 and from the RCUK Energy Programme [grant number EP/I501045]. To obtain further information on the data and models underlying this paper please contact PublicationsManager@ccfe.ac.uk. The views and opinions expressed herein do not necessarily reflect those of the European Commission.

**Figures**

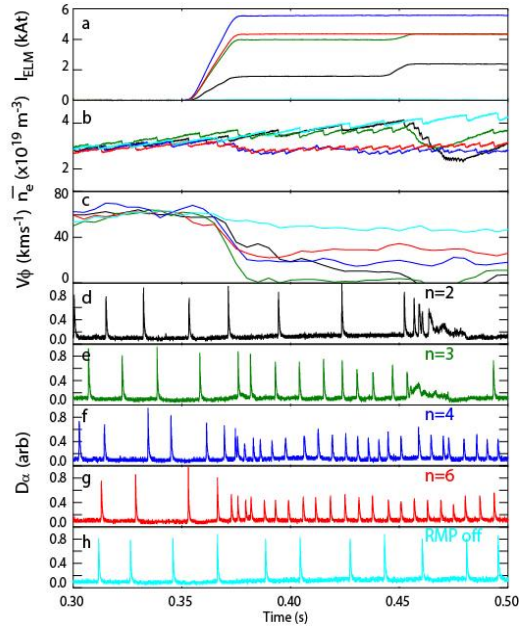

**Figure 1** For MAST: a) the current in the ELM coils ($I_{ELM}$) b) line average density, c) the toroidal rotation at the top of the pedestal ($V_\phi$) and the target $D_\alpha$ intensity for discharges with RMPs in a d) n=2, e) n=3, f) n=4, g) n=6 configurations and h) without RMPs.

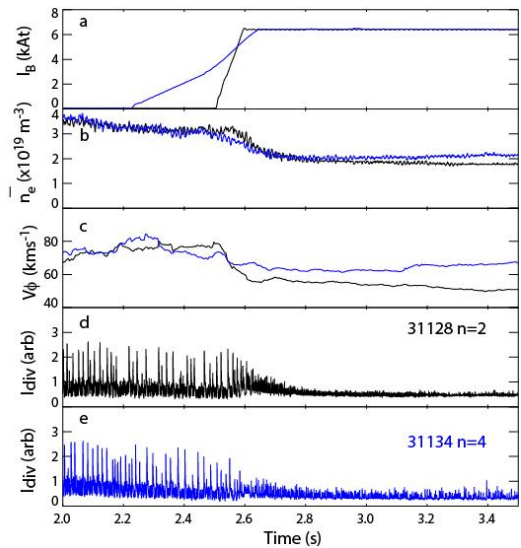

**Figure 2** For ASDEX Upgrade: a) the current in the ELM coils ($I_B$) b) line average density, c) the toroidal rotation at the top of the pedestal ($V_\phi$) and the divertor currents for discharges with RMPs in a d) n=2 and e) n=4 configuration.

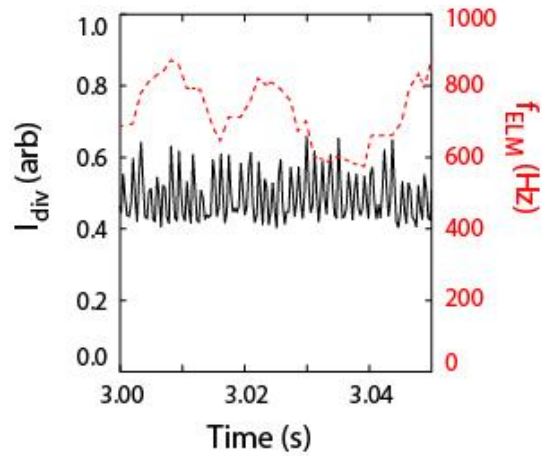

**Figure 3** For ASDEX Upgrade: The divertor currents and ELM frequency for shot 31128 during the mitigated period with RMPs in an n=2 configuration.

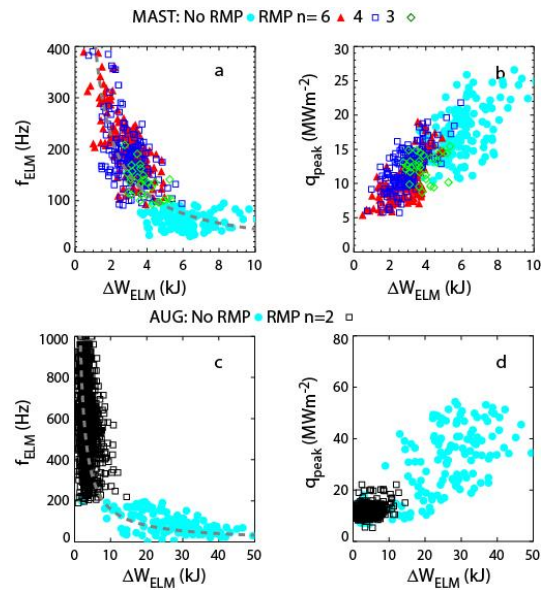

**Figure 4** a), c) ELM frequency ($f_{ELM}$) and b) d) peak divertor heat flux ($q_{peak}$) versus ELM energy loss ($\Delta W_{ELM}$) for natural and mitigated ELMs in MAST and AUG respectively.

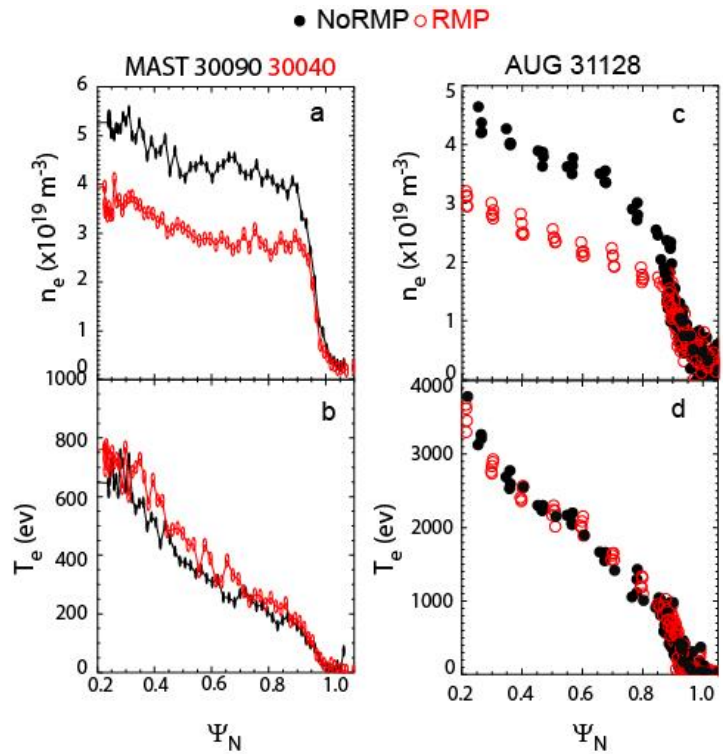

**Figure 5** The radial profiles of the electron a), c) density and b), d) temperature obtained from Thomson scattering on MAST and AUG respectively.

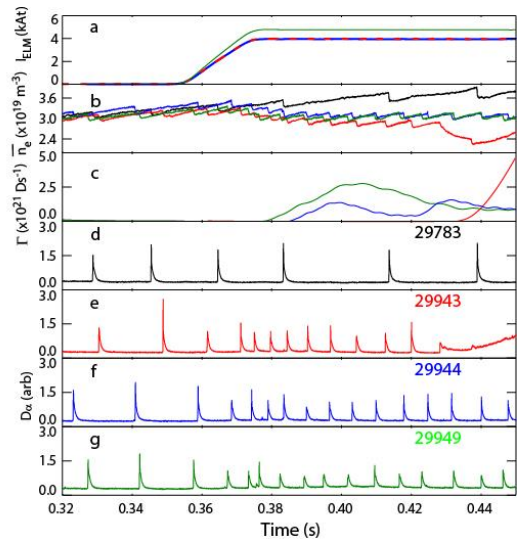

**Figure 6** For MAST: a) the current in the ELM coils ($I_{ELM}$) b) line average density, c) the gas puff fuelling rate   and the target $D_\alpha$ intensity for discharges d) without and e-g) with RMPs in n=6 configurations with different fuelling rates.

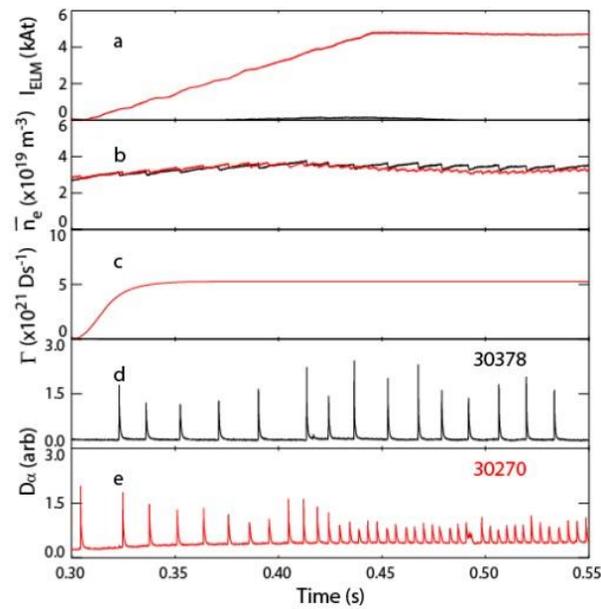

**Figure 7** For MAST: a) the current in the ELM coils ($I_{ELM}$) b) line average density, c) the gas fuelling rate and the target $D_\alpha$ intensity for discharges d) without and e) with RMPs in an n=6 configuration.

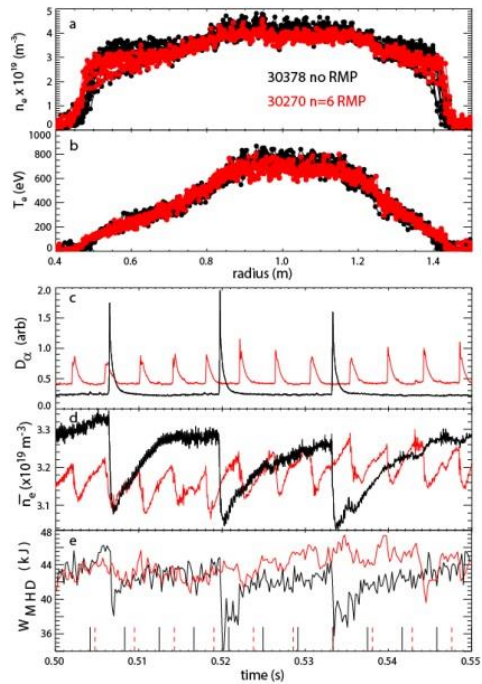

**Figure 8** For MAST: The Thomson scattering a) density and b) temperature radial profiles, c) the target $D_\alpha$ intensity d) the line average density and e) the plasma stored energy for discharges without and with RMPs in an n=6 configuration.

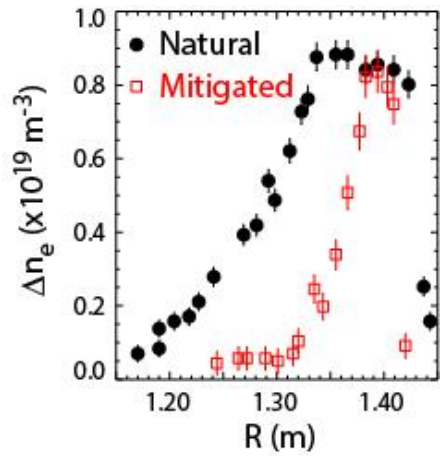

**Figure 9** For MAST: Change in the electron density profiles before and after a natural (circles) and mitigated (squares) ELM.

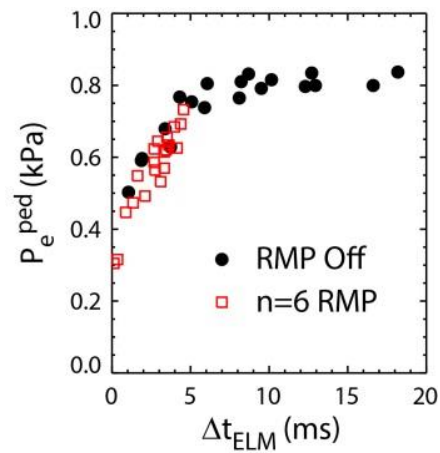

**Figure 10** Evolution of the electron pressure pedestal height during the ELM cycle for shots without (circles) and with (squares) RMPs in an n=6 configuration on MAST.

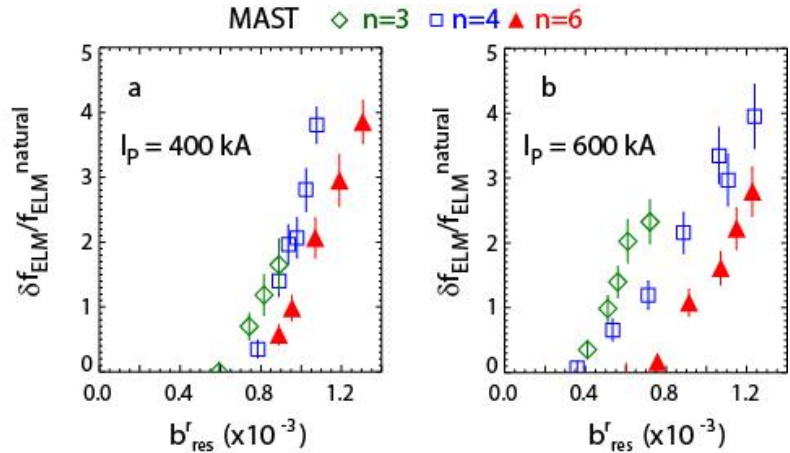

**Figure 11** Results from MAST of a scan of $I_{ELM}$ with different $n_{RMP}$: Increase in ELM frequency ($\delta f_{ELM} = f_{ELM}^{mitigated} - f_{ELM}^{natural}$) normalised to the natural ELM frequency ($f_{ELM}$) as a function of the maximum resonant component of the applied field ($b_{res}^r$) calculated in the vacuum approximation for discharges with plasma current ($I_P$) of a) 400 kA and b) 600 kA.

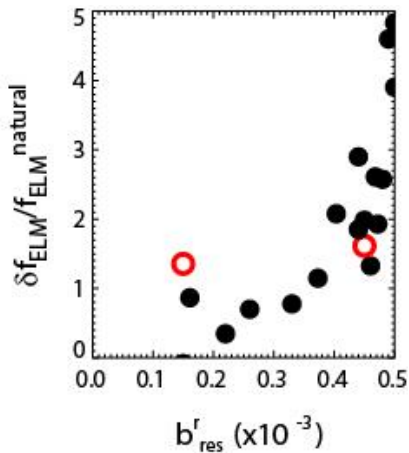

**Figure 12** Results from AUG: Normalised increase in ELM frequency ($\delta f_{ELM}/f_{ELM}^{natural}$) as a function of the maximum resonant component of the applied field ($b_{res}^r$) calculated in the vacuum approximation for discharges the RMPs in an $n_{RMP} = 2$ configuration. The open red circles are from two identical discharges which differ only in the alignment of the applied perturbation with the equilibrium field.

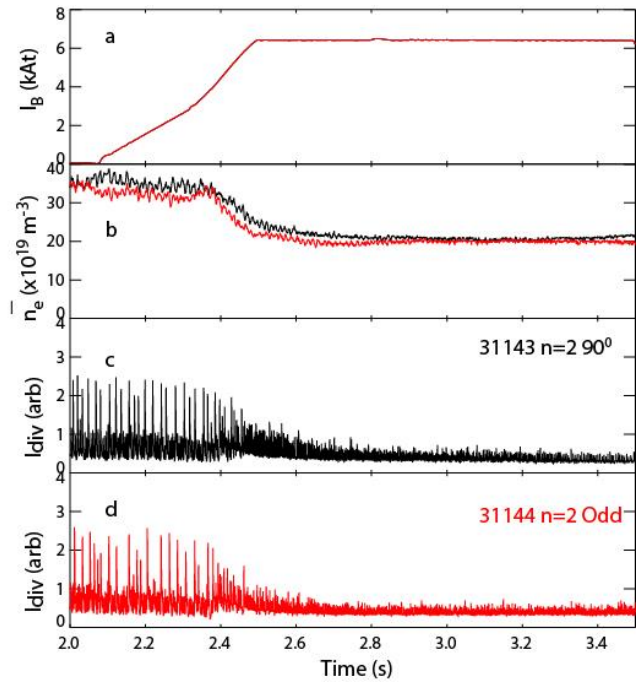

**Figure 13** For AUG: a) the current in the B-coils ($I_B$) b) line average density and the target shunt currents for discharges with RMPs in n=2 configurations with c) 90 degrees phasing and d) odd parity between upper and lower coils.

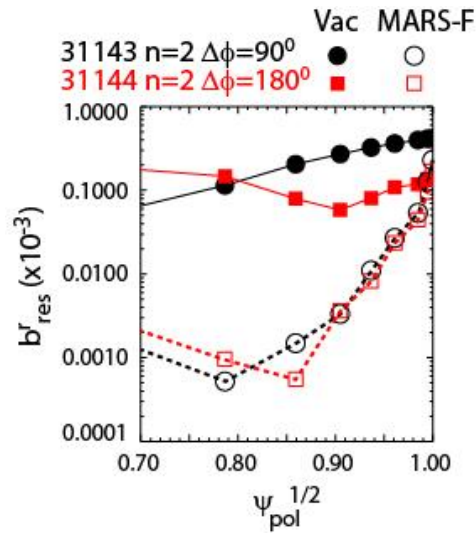

**Figure 14** Calculations of the normalised resonant component of the applied field ($b^r_{res}$) for AUG shots with the RMPs in an n=2 configuration with a toroidal phase of 90 and 180° between the coils in the upper and lower row for vacuum (solid) and including the plasma response (open).

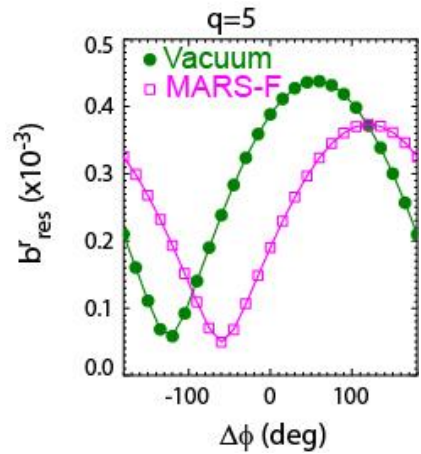

**Figure 15** The normalised resonant component of the applied field ($b^r_{res}$) at the q=5 surface corresponding to AUG shots 31143 and 31134 with the RMPs in an n=2 configuration as a function of the toroidal phase ($\Delta\phi$) between the upper and lower row of coils for vacuum (circles) and including the plasma response (squares).

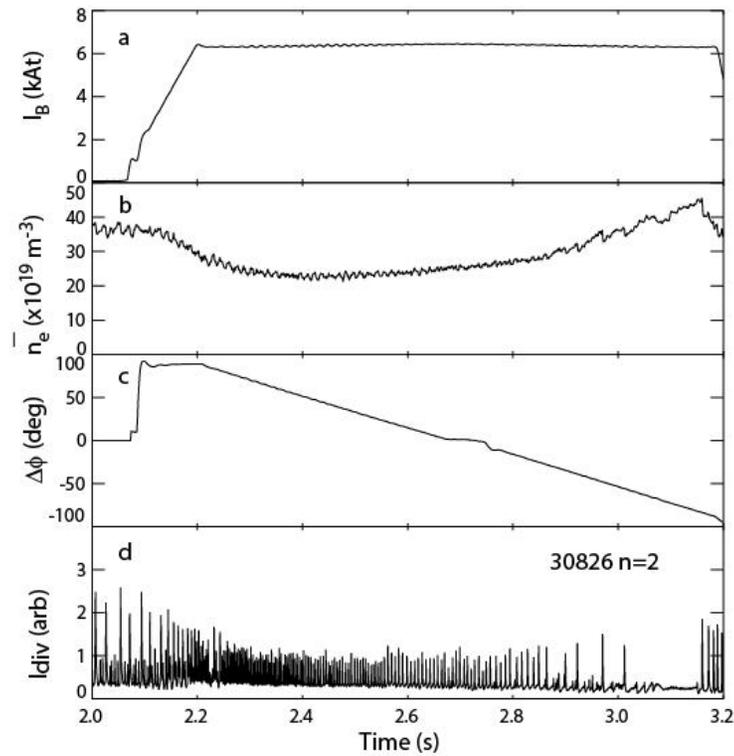

**Figure 16** For AUG: a) the current in the B-coils ($I_B$) b) line average density c) the angle between the current in the upper and lower row of coils ($\Delta\phi$) and d) the target shunt currents for discharges with RMPs in an n=2 configuration.

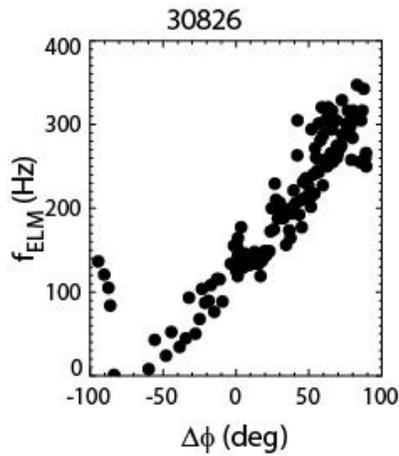

**Figure 17** For AUG shot 30826 the ELM frequency ($f_{ELM}$) versus the phasing between the current in the upper and lower row of coils ($\Delta\phi$).

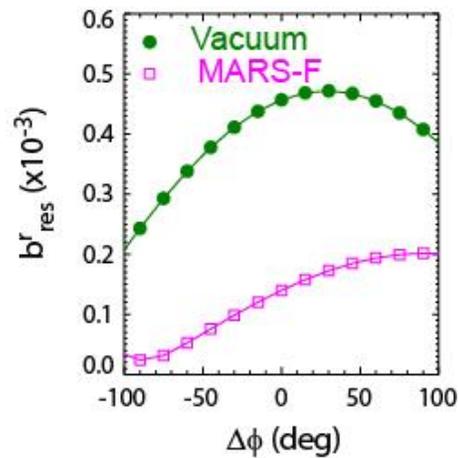

**Figure 18** The normalised resonant component of the applied field ($b^r_{res}$) at the q=5 surface for AUG shot 30826 with the RMPs in an n=2 configuration as a function of the toroidal phase ($\Delta\phi$) between the upper and lower row of coils calculated in the vacuum approximation (circles) and including the plasma response (squares).

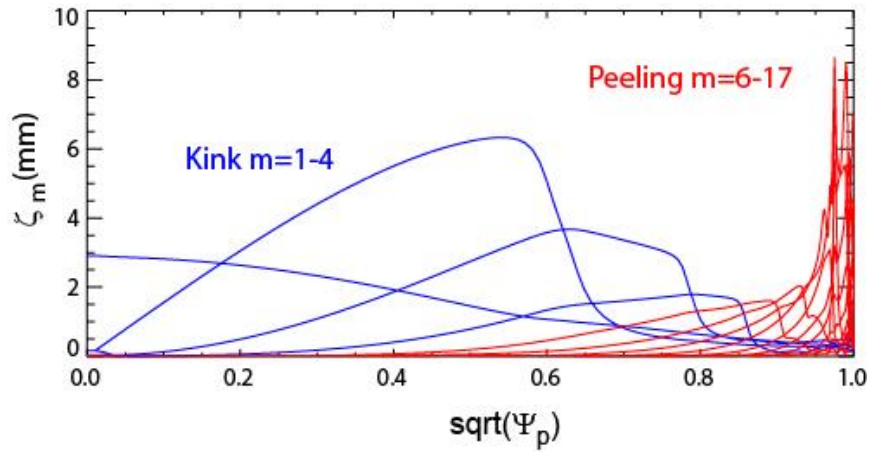

**Figure 19** An example of the radial profiles of the poloidal Fourier harmonics of the plasma displacement.

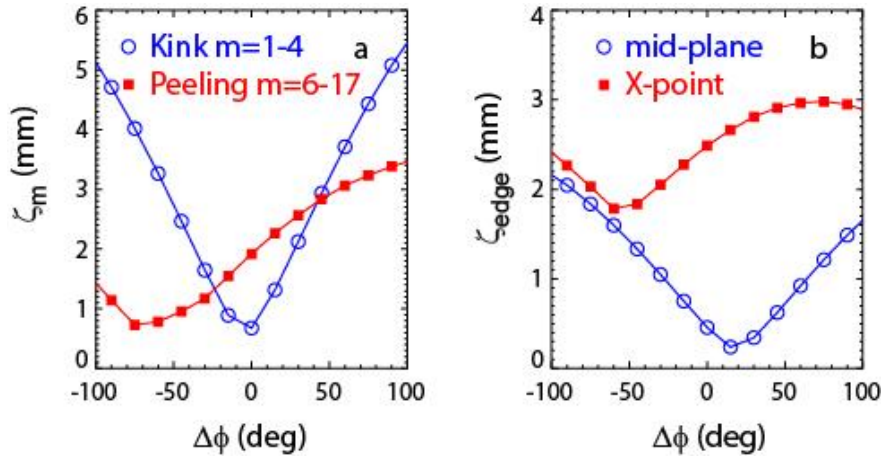

**Figure 20** Modelling for AUG shot 30826 with the RMPs in an n=2 configuration as a function of the toroidal phase (Δφ) between the upper and lower row of coils a) the maximum plasma displacement normal to the flux surfaces for the low (open circles) and high (squares) poloidal harmonics (m) and b) the plasma surface displacement at the mid-plane (circles) and X-point (squares).

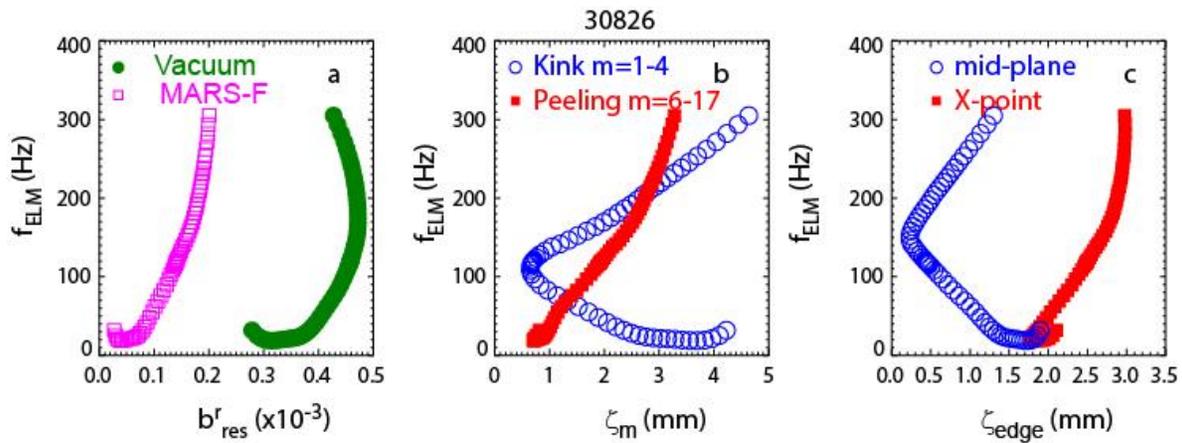

**Figure 21** Measured ELM frequency ($f_{ELM}$) for AUG shot 30826 with the RMPs in an n=2 configuration produced using a $\Delta\phi$ scan versus a) the normalised resonant component of the applied field ($b^r_{res}$) at the q=5 surface for vacuum (circles) and including the plasma response (squares), b) the maximum plasma displacement normal to the flux surfaces for the low(open circles) and high (squares) poloidal harmonics (m) and c) the plasma surface displacement at the mid-plane (circles) and X-point (squares).

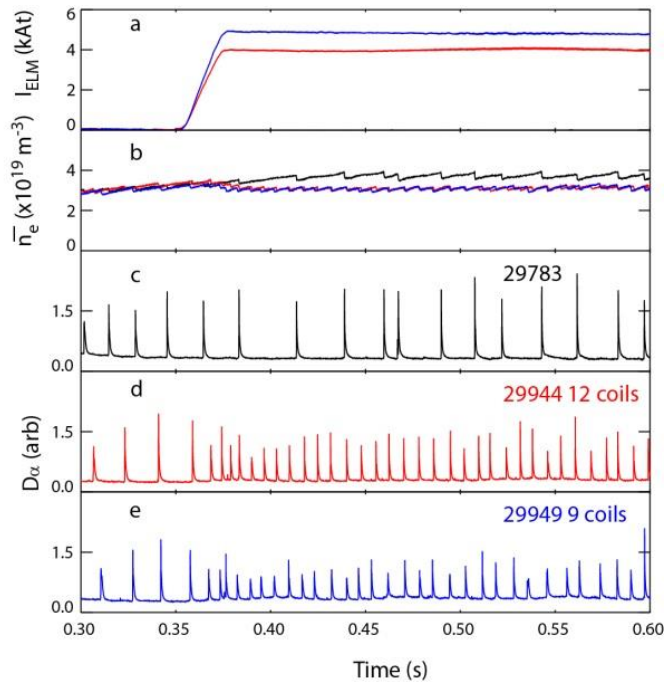

**Figure 22** For MAST: a) the current in the ELM coils (IELM) b) line average density and the target $D_\alpha$ intensity for discharges c) without RMPs and with RMPs in an $n_{RMP}$=6 configuration using  d) 12 coils and e) 9 coils.

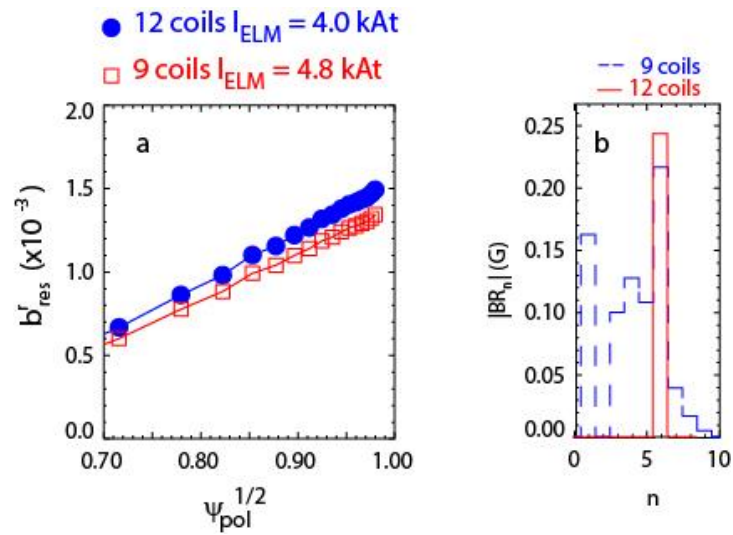

**Figure 23** a) Calculated profiles using the vacuum approximation of the normalised resonant component of the applied field ($b^r_{res}$) produced with the ELM coils in an n=6 configuration with 4.0 kAt in all 12 coils (circles) and 4.8 kAt in 9 coils (squares). b) The radial field for each toroidal harmonic of the applied field calculated at the last closed flux surface at low field side mid-plane.

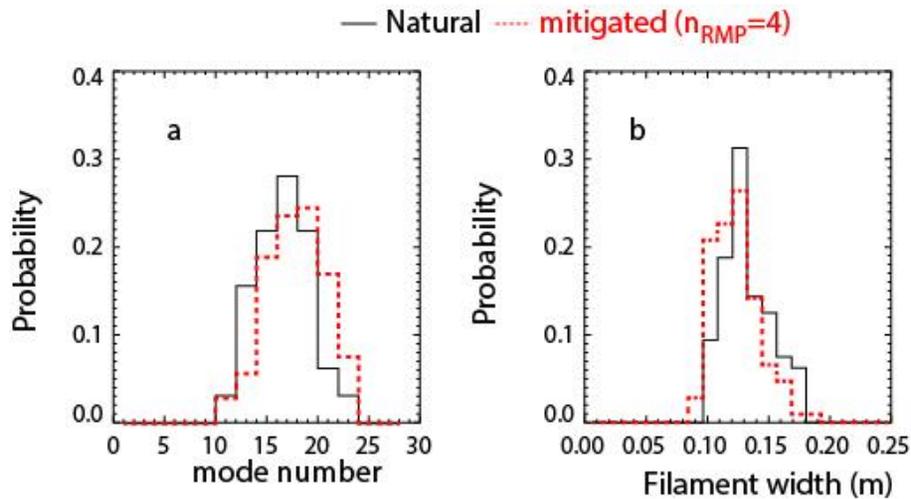

**Figure 24** Probability distribution of a) the toroidal mode number and b) the filament width for natural ELMs (solid) and ELMs mitigated (dashed) using RMPs in an $n_{RMP}$=4 configuration on MAST.

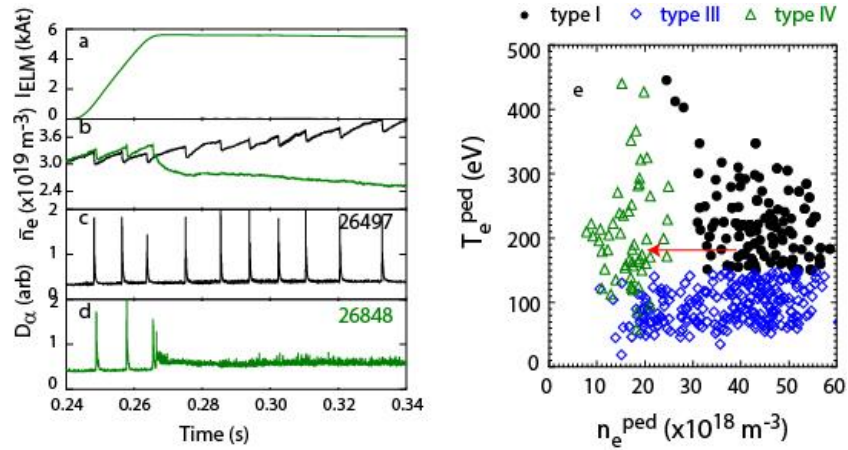

**Figure 25** For MAST: a) Time traces for a scenario 4 shot of a) coil current ($I_{ELM}$) b) line average density ($\bar{n_e}$), and $D_\alpha$ traces for shots c) without RMPs and d) with RMPs in an n=3 configuration. e) Electron temperature height versus density pedestal height as a function of ELM type from profiles obtained in the last 10 % of the ELM cycle. The arrow shows the change in pedestal characteristics from the natural to mitigated stage of d)

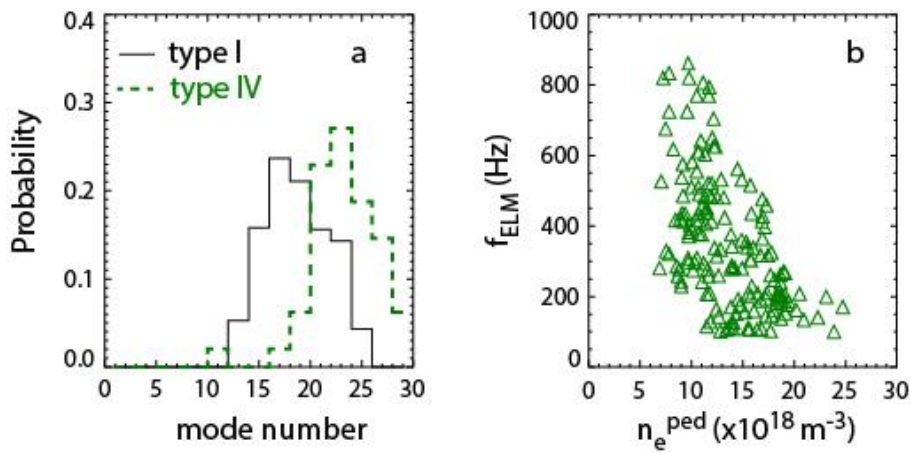

**Figure 26** For MAST a) Probability distribution of the toroidal mode number for natural ELMs (solid) and type IV ELMs (dashed). b) ELM frequency versus pedestal density for type IV ELMs.

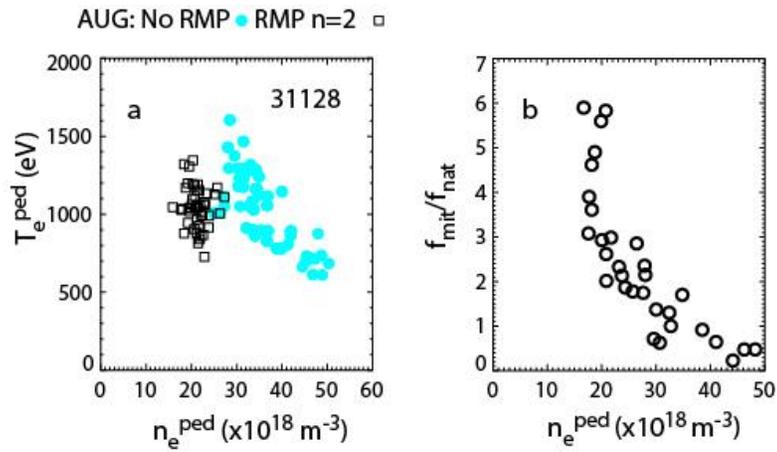

**Figure 27** For AUG a) Electron temperature pedestal height versus density pedestal height for natural (circle) and mitigated (squares) ELMs. b) Increase in ELM frequency versus density pedestal after any density pump-out.

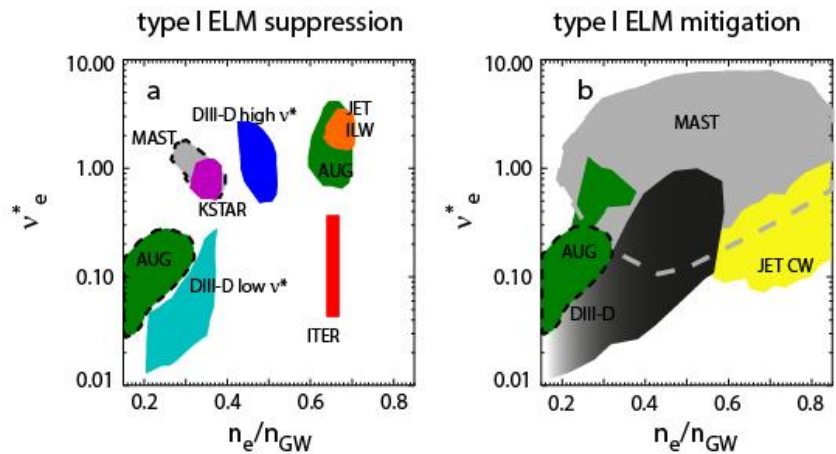

**Figure 28** Experimentally determined access condition in terms of pedestal collisionality ($\nu^*_e$) versus pedestal density as a fraction of the Greenwald density ($n_e/n_{GW}$) for a) suppression of type I ELMs and b) type I ELM mitigation. The dashed curves represent the new data shown in this paper.